# Some Relationships for the Inter-Atomic Potential Models Describing the Interaction between the Chemically Different Type Atoms

A.A. Likhachev, Yu.N. Koval, T.G. Sych and V.A. Tatarenko

*Institute for Metal Physics, National Academy of Sciences, 36 Vernadsky Str., 03680,Kiev, Ukraine,* e-mail: alexl@imp.kiev.ua

**Abstract**

Three-parametric Lenard-Jones and Morse interatomic potentials are the simplest ones, which that can be used to obtain thermophysical properties of the liquid and solid substances. Upon adjusting the model parameters to real substance properties, the interatomic potentials can be used to describe simple mono-component substance with good accuracy. Usually, these tree parameters can be found from the cohesion energy, bulk moduli and the molar volume data or the lattice parameters, obtained experimentally for pure chemically crystalline materials. In our paper, in case of chemically different atoms, for both the Lenard-Jones potential type and the Morse ones, or any other three-parametric potential type, we propose some convenient model relationships expressing the corresponding three parameters through the previously found ones for pure chemical elements.

**Introduction**

Finding the semi-empirical interatomic interaction potentials is an important task for many aims. First, it may be initial starting point for the modeling of different thermodynamic properties of liquids and solids consisting of pure chemical element and also of multiple chemical components.
The Lenard-Jones [1-3] and Morse [6] potentials are the simplest ones, which that can be used to obtain thermophysical properties of the liquid and solid substances. These potentials are mathematically simple and, therefore, often used in different computer simulation studies. Due to their mathematical simplicity and generic modeling capabilities, the mentioned potentials are probably still the most frequently studied model potentials. The Lenard-Jones potential is usually the standard choice for the development of theories for matter (especially soft-matter) as well as for the development and testing of computational methods and algorithms. Upon adjusting the model parameters to real substance properties, the interatomic potentials can be used to describe simple substance with good accuracy. Usually, these parameters can be found from the cohesion energy, bulk moduli and the molar volume data or the lattice parameters, obtained experimentally for pure chemically mono-component crystalline materials [4,5,7]. These ideas have been also applied to study some other properties of crystalline solids [8,9,11]. Nevertheless, finding similar parameters for interatomic potentials in case of chemically different interacting atoms is not so easy and the problem is still very actual.

In our paper, in case of chemically different atoms, for both the Lenard-Jones potential type and the Morse ones, or any other three-parametric potential type, we propose some convenient model relationships expressing the corresponding three parameters through the previously found ones for pure chemical elements.

## 1. Parameterizations of the Lenard-Jones and Morse Potentials

Here, we introduce two three-parametric potentials, namely, the Lenard-Jones type (LJ-type) and also the Morse type (M-type) potential interaction energy between two neighbouring same atoms in the following form:

$$u_{LJ}(r) = \varepsilon\left((r/a)^{-2n} - 2(r/a)^{-n}\right) \quad and \quad u_M(r) = \varepsilon\left(e^{-2\beta((r/a)-1)} - 2e^{-\beta((r/a)-1)}\right). \tag{1}$$

They have three parameters, LJ$\{\varepsilon, a, n\}$ and M$\{\varepsilon, a, \beta\}$, respectively. Here, r means the interatomic distance. Both interaction energy models $u_P(r)$ (with P$\{\varepsilon, a, n\}$ for LJ or P$\{\varepsilon, a, \beta\}$ for M) have minima at $r = a$, where $u_P(a) = -\varepsilon$ characterizes the bonding energy of two same atoms at the equilibrium distance between them. The exponent parameters, $n, \beta$ and $2n, 2\beta$, characterize the decrease or increase rate of the attractive energy parts and the repulsive ones, as the interatomic separation distance $r$ changes.

Instead of these three parameters, we can introduce a new set of universal parameters like the *equilibrium interatomic bond length **a**, elastic stiffness of interatomic bonds **k*** and the *elastic cutoff distance **c***, where

$$k = u_P''(a), \qquad c = \sqrt{2\varepsilon/k}. \tag{2}$$

In particular, for the Lenard-Jones potential,

$$k_{LJ} = 2\varepsilon n^2 / a^2, \qquad c = a/n. \tag{3}$$

And, for the Morse model, we obtain a similar result:

$$k_M = 2\varepsilon\beta^2 / a^2, \qquad c = a/\beta. \tag{4}$$

Both interatomic potentials of LJ and M types and as well as so-called Elastic-Bond potential model calculated at the $\beta = n = 5$ values are shown in Fig.1. In these both cases and also in any other three-parametric interatomic potential models we will use the following 'model-independent' (universal) relationships:

$$k = 2\varepsilon/c^2, \qquad \varepsilon = \frac{1}{2}kc^2. \tag{5}$$

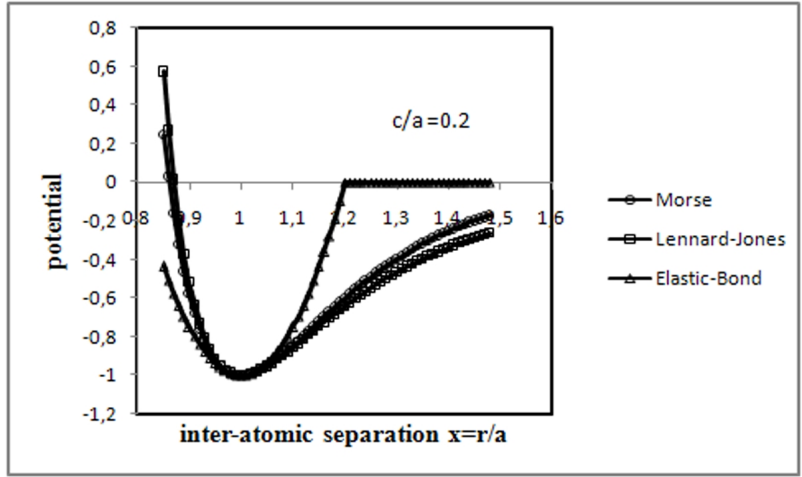

**Fig.1. Three-parametric Lenard-Jones, Morse and parabolic Elastic-Bond interatomic potentials calculated at equal bonding energy $\varepsilon = 1$ and $\beta = n = 5$.**

Here, we have introduced so called Elastic-Bond potential:

$$u_{EB}(r) = \varepsilon\left(-1 + (\gamma(r/a - 1))^2\right) \quad (6)$$

and the equations similar to Eqs.(3), (4), obtained for the Lenard-Jones and Morse potentials,

$$k_{EB} = 2\varepsilon\gamma^2/a^2, \qquad c = a/\gamma. \quad (7)$$

It also shown in Fig.1 and calculated at $\gamma = \beta = n = 5$. As it follows from Fig.1, the elastic cutoff parameter $c$ introduced in Eq.2 characterizes interatomic distance where Elastic-Bond potential takes a zero value: $u_{EB}(a + c) = 0$ that means that the elastic bond is broken at this distance. So, this parameter can be used for any interatomic potential model as well.

*Summary_1.*

In present subsection, for any three-parametric potential $u_P(r)$, instead of corresponding three parameters we can introduce a new set of universal parameters like the **equilibrium interatomic bond length** $a$, where: $u'_P(a) = 0$, the **inter-atomic coupling energy** $u_P(a) = -\varepsilon$, *the elastic stiffness of inter-atomic bond* $k = u''_P(a)$ and the **elastic cutoff distance** $c = \sqrt{2\varepsilon/k}$. Finally, we have introduced convenient dimensionless potential parameter $\eta = a/c$. Using these ones, the inter-atomic potential itself can be represented as follows:

$$u_P(r) = \varepsilon\varphi_\eta\left(\frac{r}{a}\right). \quad (8)$$

Here, the dimensionless potential function $\varphi_\eta(x)$ depends only on the dimensionless interatomic separation distance $x$ and a single potential parameter $\eta$. In case of the Lenard-Jones and Morse potentials: $\eta = \beta = n$.

2. **Finding parameters from the experimental data**

To find these parameters most of researchers use the cohesion energy, bulk moduli and the molar volume (or lattice parameter) data such as represented in Table1 for some f.c.c and b.c.c transition metals taken from [10, 12-14].

*Table1. Cohesive energy, bulk modulus and molar volume interpolated to $0K^0$ temperature for some f.c.c and b.c.c metals (experimental data).*

| Units | V b.c.c | Nb b.c.c | Ta b.c.c | Cr b.c.c | Mo b.c.c | W b.c.c | Fe b.c.c | |
|---|---|---|---|---|---|---|---|---|
| $E_m$(kJ/mol) | 512 | 730 | 782 | 395 | 658 | 859 | 413 | |
| K(GPa) | 160 | 170 | 200 | 160 | 230 | 310 | 170 | |
| $V_m$($10^{-6}m^3$) | 8.337 | 10.84 | 10.87 | 7.232 | 9.334 | 9.550 | 7.092 | |
| $KV_m/E_m$ | 2.59 | 2.52 | 2.78 | 2.93 | 3.26 | 3.45 | 2.92 | |
| | Ni | Pd | Pt | Cu | Ag | Au | Rh | Ir |
| Units | f.c.c | f.c.c | f.c.c | f.c.c | f.c.c | f.c.c | f.c.c | f.c.c |
| $E_m$(kJ/mol) | 428 | 376 | 564 | 336 | 284 | 368 | 554 | 670 |
| K(GPa) | 180 | 180 | 230 | 140 | 100 | 220 | 380 | 320 |
| $V_m$($10^{-6}m^3$) | 6.589 | 8.851 | 9.095 | 7.092 | 10.28 | 10.21 | 8.266 | 8.520 |
| $KV_m/E_m$ | 2.77 | 4.23 | 3.71 | 2.95 | 3.61 | 6.32 | 5.47 | 3.42 |

Using Eq.(1), one can write the molar internal energy at zero temperature for the crystalline material as function of the nearest neighbour distance *r* as follows:

$$U(r) = \frac{1}{2} N_A \sum_{i=1}^{K_c} z_i u_P(rp_i) = \frac{1}{2} N_A \varepsilon \sum_{i=1}^{K_c} z_i \varphi_\eta \left(\frac{r}{a} p_i\right) = \frac{1}{2} N_A \varepsilon \Phi_\eta \left(\frac{r}{a}\right) \qquad (9)$$

Here, summation is performed over a set of *Kc* nearest coordination spheres: $i = 1, 2, ... K_c$, where $z_i$ denotes the number of atoms on the *i-th* coordination sphere. $N_A = 6.02214076 \times 10^{23}$/mol is the well-known Avogadro's number, and $p_i$ is a set of dimensionless distances to the *i-th* coordination sphere. We also, introduced a dimensionless Lenard-Jones or Morse potential function $\varphi_\eta(x)$ having a minimum at *x* = 1, where $\varphi_\eta(1) = -1$ and the second derivative value is $\varphi_\eta''(1) = 2\eta^2$ and $\eta = \beta = n$ for M and LD type potentials, respectively. Further, we will denote $\Psi_\eta(x) = \frac{1}{2}\Phi_\eta\left(\frac{r}{a}\right) = \frac{1}{2}\Phi_\eta(x)$, as the dimensionless crystal energy per atom. Normally, the internal energy of a crystal must be represented as a function of the molar volume $Vm = N_A v(r)$ where, *v(r)* is the volume per single atom in a given crystalline material. For many ideal crystal structures like f.c.c, h.c.p and b.c.c there is a simple relationship between the atomic volume and the nearest neighbour distance $v(r) = qr^3$ where, *q* is a constant dependent on a particular crystal geometry. At the equilibrium distance $r = r_e$, where,

$$\left(\frac{d}{dr} U(V(r))\right)_{r=r_e} = N_A \varepsilon \Psi_\eta'(x_e) = 0, \qquad (10)$$

we can obtain a dimensionless ratio $x_e = r_e / a$ and also the equilibrium molar volume:

$$V_m(r_e) = N_A q r_e^3. \tag{11}$$

Therefore, using Eqs.(8-9), one can obtain the bulk modulus *K* at zero temperature as follows:

$$K = V_m\left(\frac{d^2U(V)}{dV^2}\right)_{V=V_m} = \varepsilon \frac{1}{9qr_e a^2}\Psi_\eta''(x_e). \tag{12}$$

On the other hand, at zero temperature, the molar cohesive energy is:

$$E_m = -U(V_m) = -N_A \varepsilon \Psi_\eta(x_e). \tag{13}$$

From Eqs.(12), (13) it follows an important dimensionless relationship between the cohesive energy, bulk modulus and the molar volume:

$$\frac{V_m K}{E_m} = -\frac{1}{9} x_e^2 \Psi_\eta''(x_e)/\Psi_\eta(x_e) \tag{14}$$

Here, $x_e = x_e(\eta)$ is a value minimizing the dimensionless cohesion energy function that can be obtained solving Eq.(10). It depends obviously on the dimensionless parameter $\eta = \{n, \beta, \gamma\}$ of any interatomic potential discussed in the present paper or some other ones too. So, Eq.(14) gives a possibility finding this one parameter as function of the experimental dimensionless quantity $(V_m K / E_m)$:

$$\eta = \eta(V_m K / E_m). \tag{15}$$

The second parameter, namely, the bonding energy one can find from Eq.(13):

$$\varepsilon = -\frac{E_m / N_A}{\Psi_\eta(x_e(\eta))}. \tag{16}$$

And the final, third parameter, i.e, the bonding length *a* is

$$a = d / x_e(\eta). \tag{17}$$

Here, *d* is the experimentally measured nearest neighbour distance in b.c.c, f.c.c, or h.c.p structure or calculated from the known atomic volume:

$$v(d) = qd^3 = V_m / N_A, \tag{18}$$

where, $q_{bcc} = 4/3\sqrt{3} = 0.7698;\quad q_{fcc} = 1/\sqrt{2} = 0.7071$, and, obviously,

$$d_{bcc} = \sqrt{3}(2v_{bcc})^{1/3}/2; \quad d_{fcc} = (4v_{fcc})^{1/3}/\sqrt{2}. \tag{19}$$

The data given in Table1 can be used to calculate all the interatomic potential parameters. Normally, it can be done applying appropriate numerical methods. But here, we restrict ourselves by the simplest nearest neighbour approach by setting in Eq.(9) *Kc = 1* and $\Psi_\eta(x) = \frac{z_1}{2}\varphi_\eta(x)$ and finding from Eq.(10)

$\varphi'(x_e) = 0 \Rightarrow x_e = 1$. After that, one can see: $\Psi_\eta(1) = -\dfrac{z_1}{2}$  $\Psi_\eta''(1) = \dfrac{z_1}{2} 2\eta^2$ ($\eta = \{n, \beta, \gamma\}$). Finally, using Eq.(14) we obtain:

$$\frac{V_m K}{E_m} = \frac{2}{9}\eta^2 \quad \text{or} \quad \eta = 3\sqrt{V_m K / 2E_m}. \tag{20}$$

At the same time, from the Eqs.(16), (17) we find the two remained potential parameters:

$$\varepsilon = \frac{2}{z_1}(E_m / N_A); \quad \text{and} \quad a = d. \tag{21}$$

All the parameters characterizing the model Lenard-Jones and Morse potentials consistent with experimental data are collected in Tables 2a-2b. It should mentioned that the potential parameters for the b.c.c lattice were calculated using both the first and second coordination sphere setting in Eq.(9) **Kc = 2**. The reason of that is the known elastic instability of b.c.c lattices in the nearest neighbour approach in respect to it's shear distortion.

*Table2a. Cohesive energy per atom (eV), volume per atom ($A^3$) and the nearest neighbour distance (A) together with the potential parameters calculated from the experimental data in the nearest neighbour approach for some bcc metals.*

| BCC Metal | $\varepsilon_a$ (eV) | $v_a$ ($A^3$) | d (A) | β | ε (eV) | k (eV/$A^3$) | c (A) |
|---|---|---|---|---|---|---|---|
| V  | 5,31 | 13,84401 | 2,619959 | 3,413942 | 0,884493 | 3,003638 | 0,767429 |
| Nb | 7,57 | 18,00037 | 2,859572 | 3,367492 | 1,261094 | 3,497746 | 0,84917  |
| Ta | 8,11 | 18,05018 | 2,862208 | 3,536948 | 1,350925 | 4,125877 | 0,809231 |
| Cr | 4,09 | 12,00578 | 2,49845  | 3,631116 | 0,682373 | 2,882639 | 0,688067 |
| Mo | 6,82 | 15,49293 | 2,720102 | 3,830144 | 1,136712 | 4,507546 | 0,710183 |
| W  | 8,90 | 15,85826 | 2,741316 | 3,940178 | 1,483945 | 6,131421 | 0,695734 |
| Fe | 4,28 | 11,7733  | 2,482219 | 3,624914 | 0,713468 | 3,043125 | 0,684766 |

*Table2b. Cohesive energy per atom (eV), volume per atom ($A^3$) and the nearest neighbour distance (A) together with the potential parameters calculated from the experimental data in the nearest neighbour approach for some fcc metals.*

| FCC Metal | $\varepsilon_a$ (eV) | $v_a$ ($A^3$) | d (A) | β | ε (eV) | k (eV/$A^3$) | c (A) |
|---|---|---|---|---|---|---|---|
| Ni | 4,44 | 10,94303 | 2,492016 | 3,530581 | 0,739381 | 2,968171 | 0,705837 |
| Pd | 3,90 | 14,69587 | 2,749393 | 4,362912 | 0,64955 | 3,27131 | 0,630174 |
| Pt | 5,85 | 15,11101 | 2,775041 | 4,085952 | 0,974325 | 4,224556 | 0,679167 |
| Cu | 3,48 | 11,7733 | 2,553511 | 3,643487 | 0,580449 | 2,363485 | 0,700843 |
| Ag | 2,94 | 17,07046 | 2,890148 | 4,030509 | 0,490617 | 1,908326 | 0,717068 |
| Au | 3,81 | 16,95422 | 2,883573 | 5,332917 | 0,63573 | 4,348807 | 0,540712 |
| Rh | 5,74 | 13,73275 | 2,687969 | 4,961351 | 0,957049 | 6,521024 | 0,541782 |
| Ir | 6,94 | 14,14789 | 2,714786 | 3,923009 | 1,157442 | 4,833891 | 0,692016 |

## *Summary_2.*

In present subsection, we have discussed a procedure of finding the potential parameters using the cohesion energy, bulk modulus and the molar volume (or lattice parameter) data represented in the Table1 for some f.c.c and b.c.c transition metals. The idea using these experimental data has been proposed quite long ago and, then, was applied by many other researchers mainly limiting themselves by LJ- and M- cases [4,5,7]. The main reason for us is to formulate the finding procedure to be maximally independent on a particular potential type. So, we did everything in dimensionless form. The basic crystal energy function we have defined as a sum of the individual dimensionless interatomic potentials taken over some appropriate number **Kc** of coordination spheres according to Eq.(9).

$$\Psi_\eta(x) = \frac{1}{2} \sum_{i=1}^{Kc} z_i \varphi_\eta(x p_i). \tag{22}$$

Then, making minimizing procedure, one can find the equilibrium dimensionless nearest-neighbour distance $x = x_e(\eta)$ dependent on the potential parameter $\eta$ in a given crystal lattice with a given set of coordination numbers $\{z_i\}$ and a set of dimensionless distances $\{p_i\}$ to the *i-th* coordination sphere. Finally, using Eq.(14), we obtain parameter $\eta$ as a function of the only dimensionless experimental parameter $\chi = V_m K / E_m : \eta = \eta(\chi)$, and also two remaining potential parameters:

$$\varepsilon(\chi) = -e_m / \Psi(x_e(\eta(\chi))) \qquad \text{and} \qquad a(\chi) = d / x_e(\eta(\chi)). \tag{23}$$

Here, $e_m = E_m / N_A$ is the cohesive energy per atom and $d$ is the experimentally measured distance between the nearest atoms in b.c.c f.c.c or h.c.p type crystals.

## 3. Modeling interatomic potentials for chemically different type atoms

In present section, we will propose some simple ideas, which allow constructing the interatomic potential for two chemically different atoms **A** and **B**, shown schematically in Fig.2.

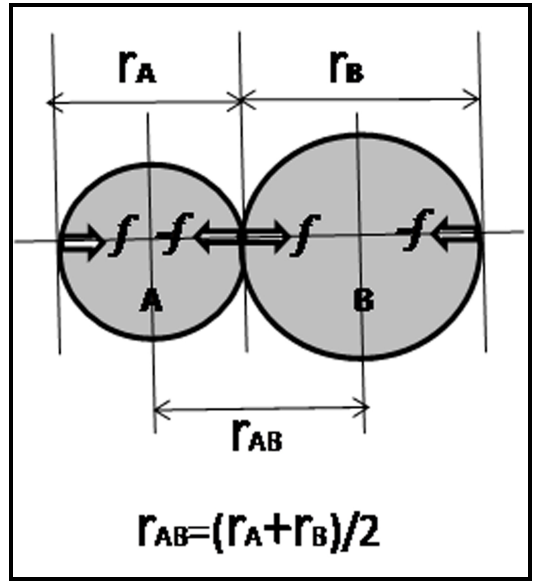

**Fig.2. This scheme represents two ball-shaped chemically different atoms A and B producing weak forces when the distance between them slightly changes.**

These atoms are placed at some interatomic distance $r_{AB} = (r_A + r_B)/2$ where, $r_A$ and $r_B$ denote the bond length for AA or BB type interacting atoms, respectively. Here $f$ is a force taking zero value if atoms A and B are at their equilibrium distances: $r_A = a_A$ and $r_B = a_B$. But the interaction forces acting between the atoms (all equal due to Newton's third law) will supposedly change linearly, if $r_A$ and $r_B$ will experience some small changes: $r_A = a_A + \delta a_A$ and $r_B = a_B + \delta a_B$, as follows from equations below.

$$f = k_A \delta a_A \quad \text{and} \quad f = k_B \delta a_B \tag{23}$$

Here, $k_A, k_B$ are the elastic bond stiffness parameters introduced in previous section. A similar relationship can be supposedly written for the AB bonds:

$$f = k_{AB} \delta a_{AB}; \quad \text{where}, \quad \delta a_{AB} = (\delta a_A + \delta a_B)/2 \tag{24}$$

Combining these equations, one can obtain relationships for the elastic-bond stiffness parameters for the AB-type interatomic bonds as function of the stiffness parameters for the AA and BB ones, respectively.

$$1/k_{AB} = (1/k_A + 1/k_B)/2 \tag{25}$$

Along with this important equation, one can use a similar additive rule for the equilibrium bond length $a$. The elastic cutoff distance $c_{AB}$ introduced in the previous section we will represent generally consisting of the half-additive contribution $(c_A + c_B)/2$ and a difference $c_{AB}^{exc} = c_{AB} - (c_A + c_B)/2$ as indicated in Eq.(26).

$$a_{AB} = (a_A + a_B)/2, \quad c_{AB} = (c_A + c_B)/2 + c_{AB}^{exc}. \tag{26}$$

Later we will show that this so-called c-excess parameter $c_{AB}^{exc}$ is small, but nevertheless plays a very important role. These three relationships give us full description of all three parameters for the Lennard-Jones and Morse potential, or any other three-parametric potentials as well. In particular, the bonding energy between the AB atoms can be easily found from the following equation:

$$c_{AB}^2 / \varepsilon_{AB} = \left( c_A^2 / \varepsilon_A + c_B^2 / \varepsilon_B \right) / 2. \tag{27}$$

One can also obtain the following relationships for the exponent parameters $n, \beta$ Lennard-Jones and Morse potentials: $n_{AB} = \beta_{AB} = a_{AB} / c_{AB}$.

Eq.(27) is very important for many applications for ordered or disordered binary alloys to predict their thermodynamic properties at different atomic concentrations.

In order to analyze and discuss this equation, it is more convenient to express it in a dimensionless form:

$$\frac{1}{(\varepsilon_{AB}/\varepsilon)} = \left[ \left( c_A^2 / c_{AB}^2 \right) \frac{1}{(\varepsilon_A / \varepsilon)} + \left( c_B^2 / c_{AB}^2 \right) \frac{1}{(\varepsilon_B / \varepsilon)} \right] / 2, \quad \text{where}, \quad \varepsilon = (\varepsilon_A + \varepsilon_B)/2. \tag{28}$$

One can also express parameters $\varepsilon_A, \varepsilon_B$ as follows:

$$\varepsilon_A = (\varepsilon_A + \varepsilon_B)/2 + (\varepsilon_A - \varepsilon_B)/2, \qquad \varepsilon_B = (\varepsilon_A + \varepsilon_B)/2 - (\varepsilon_A - \varepsilon_B)/2. \tag{29}$$

The same can be done for parameters $c_A, c_B$ as well:

$$c_A = (c_A + c_B)/2 + (c_A - c_B)/2, \qquad c_B = (c_A + c_B)/2 - (c_A - c_B)/2 \tag{30}$$

or the corresponding dimensionless relationships:

$$\varepsilon_A / \varepsilon = 1 + (\varepsilon_A - \varepsilon_B)/2\varepsilon, \qquad \varepsilon_B / \varepsilon = 1 - (\varepsilon_A - \varepsilon_B)/2\varepsilon, \tag{31}$$

and

$$c_A / c_{AB} = 1 + (c_A - c_B)/2c, \qquad c_B / c_{AB} = 1 - (c_A - c_B)/2c. \tag{32}$$

Therefore, Eq.(28) is expressed through the following two dimensionless parameters with $c = (c_A + c_B)/2$ and $\varepsilon = (\varepsilon_A + \varepsilon_B)/2$:

$$w_{AB} = (c_A - c_B)/2c \quad (|w_{AB}| \leq 1), \qquad \omega_{AB} = (\varepsilon_A - \varepsilon_B)/2\varepsilon \quad (|\omega_{AB}| \leq 1). \tag{33}$$

So, finally, from Eq.(28-33) we obtain:

$$\frac{1}{(\varepsilon_{AB}/\varepsilon)} = \left( \frac{c}{c_{AB}} \right)^2 \left( \frac{(1+w_{AB})^2}{(1+\omega_{AB})} + \frac{(1-w_{AB})^2}{(1-\omega_{AB})} \right) / 2 \tag{34}$$

and then find:

$$\frac{\varepsilon_{AB}}{\varepsilon} = 2\left(\frac{c_{AB}}{c}\right)^2 \left(\frac{(1+w_{AB})^2}{(1+\omega_{AB})} + \frac{(1-w_{AB})^2}{(1-\omega_{AB})}\right)^{-1} = \left(\frac{c_{AB}}{c}\right)^2 F_{AB}(\omega_{AB}, w_{AB}). \qquad (35)$$

One can observe that $F_{AB}(\omega_{AB}, w_{AB})$ takes exactly unit values $F_{AB}(\omega_{AB}, w_{AB}) = 1$ everywhere at the straight line $\omega_{AB} = w_{AB}$ within the square box region $-1 \leq \omega_{AB} \leq 1;\ \ and\ \ -1 \leq w_{AB} \leq 1$ and $F_{AB}(\omega_{AB}, w_{AB}) < 1$ in other cases.

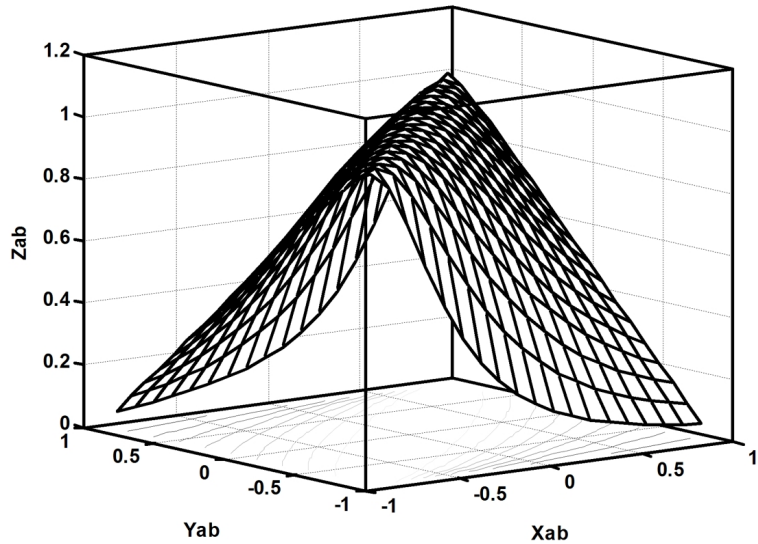

**Fig.2a, representing 3D image of the $F_{AB}(w_{AB}, \omega_{AB})$-surface calculated from Eq.35. Here, $X_{AB} = w_{AB} = (c_A - c_B)/(c_A + c_B)$, $Y_{AB} = \omega_{AB} = (\varepsilon_A - \varepsilon_B)/(\varepsilon_A + \varepsilon_B)$ and $Z_{AB} = F_{AB}(w_{AB}, \omega_{AB})$.**

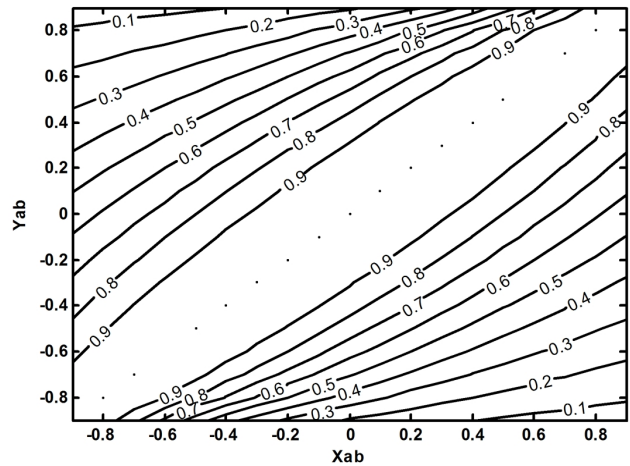

**Fig. 2b, representing 2D image of the $F_{AB}(w_{AB}, \omega_{AB})$ -surface calculated from Eq.35. Here, $X_{AB} = w_{AB} = (c_A - c_B)/(c_A + c_B)$ and $Y_{AB} = \omega_{AB} = (\varepsilon_A - \varepsilon_B)/(\varepsilon_A + \varepsilon_B)$.**

*Summary_3.*

In present section, we have proposed some simple ideas, which allowed constructing the interatomic potential parameters for two chemically different atoms **A** and **B**. For this aim, we have considered the interatomic distance changes $\delta_{AA}(f), \delta_{BB}(f), \delta_{AB}(f)$ produced by the same weak force $f$ applied to **AA**, **BB** and **AB** atomic pairs. We also assumed that, similarly to the equilibrium interatomic bond length rule $a_{AB} = (a_{AB} + a_{AB})/2$ the additional distance changes will follow the same rule: $\delta_{AB} = (\delta_{AB} + \delta_{AB})/2$. As a result, the important relationship between the interatomic stiffness parameters, $1/k_{AB} = (1/k_A + 1/k_B)/2$ has been obtained in Eq.(25). It also gives a corresponding relationship for the energy coupling parameters: $c_{AB}^2/\varepsilon_{AB} = (c_A^2/\varepsilon_A + c_B^2/\varepsilon_B)/2$. Finally' introducing two new variables $w_{AB} = (c_A - c_B)/(c_A + c_B)$ and $\omega_{AB} = (\varepsilon_A - \varepsilon_B)/(\varepsilon_A + \varepsilon_B)$, one can obtain a fully dimensionless relationship

$$\varepsilon_{AB}/\varepsilon = (c_{AB}/c)^2 F_{AB}(\omega_{AB}, w_{AB}). \tag{36}$$

where $c = (c_A + c_B)/2$, $\varepsilon = (\varepsilon_A + \varepsilon_B)/2$, and $F_{AB}(\omega_{AB}, w_{AB})$ is the function defined in Eq.35. In present subsection in Eq.(26), we have also assumed that $c_{AB} = c + c_{AB}^{exc}$ and obtained Eq.(35). There is a serious reason to introduce a new positively defined function $g_{AB}(\omega_{AB}, w_{AB}) = (c_{AB}/c)^2 = (1 + c_{AB}^{exc}/c)^2 \approx 1 + 2c_{AB}^{exc}/c$, which depends on the same variables like $F_{AB}(\omega_{AB}, w_{AB})$ does. As $F_{AB}(\omega_{AB}, w_{AB})$, and $g_{AB}(\omega_{AB}, w_{AB})$ must approach to the unit value as $(\omega_{AB} \to 0, w_{AB} \to 0)$ and both can be expanded as $F_{AB}(\omega_{AB}, w_{AB}) \approx 1 - (\omega_{AB} - w_{AB})^2$ and $g_{AB}(\omega_{AB}, w_{AB}) \approx 1 + (P1(\omega_{AB})^2 + P2\omega_{AB}w_{AB} + P3(w_{AB})^2)$ in this limit respectively. The fact, that these expansions do not contain the linear terms is due to the inversion symmetry requirement $g_{AB}(\omega_{AB}, w_{AB}) = g_{AB}(-\omega_{AB}, -w_{AB})$. So, in this case we will have:

$$\varepsilon_{AB}/\varepsilon = \approx 1 + ((P1-1)(\omega_{AB})^2 + P2(\omega_{AB})(w_{AB}) + P3(w_{AB})^2). \tag{37}$$

Here *P1*, *P2* and *P3* are some fitting constants, which should be found from the analysis of some (minimum three) binary phase diagrams.

We are not going to discuss the appropriate procedure for that in details in the present paper and will do that somewhere else. We should only to mention that it is convenient to choose the disordered binary alloys like Cu-Ni, Ni-Au and Cu-Rh. All them have a good solubility at high temperatures, but, experience decomposition below some characteristic temperatures $T_{max}$ and consentrations $x_{max}$. In case of Cu-Ni these parameters are ($t_{max}$=355C$^0$, $x_{max}$=0.673) for Ni-Au ($t_{max}$=812C$^0$, $x_{max}$=0.29) and for Cu-Rh alloy ($t_{max}$=812C$^0$, $x_{max}$=0.29) [15] correspondingly. There is also a simple relationship between these parameters the energy excess energy $(1/x_{max} + 1/(1-x_{max}))(kT_{max}/Z\varepsilon^{exc})=2$, where $T_{max}$ is the absolute temperature in K$^0$ and Z=12 is the coordination number in our case of f.c.c.-lattice.